\documentclass[11pt,a4paper]{article}
\usepackage{amsmath,amssymb,titling,authblk}
\usepackage[top=2.3cm,right=2.3cm,left=2.3cm,bottom=2.3cm]{geometry}
\usepackage{graphicx}
\usepackage{color} 
\usepackage{slashed}
\usepackage[pdfstartview=FitH,colorlinks=true,linkcolor=blue,anchorcolor=red,citecolor=magenta,urlcolor=blue]{hyperref}
\usepackage{amscd}
\usepackage[normalem]{ulem}
\usepackage{appendix}
\usepackage{bbold}
\usepackage{mathrsfs}
\usepackage{pdfsync}
\usepackage{bbm}
\usepackage{bm}
\usepackage[arrow,matrix,curve]{xy}
\usepackage{bbding}
\usepackage{wasysym}
\usepackage{booktabs}
\usepackage{siunitx}

\usepackage{epsf}
\usepackage{epsfig}
\usepackage{wrapfig}
\usepackage[utf8]{inputenc}
\usepackage{subcaption}
\usepackage{cite}
\DeclareCaptionFormat{custom}
{%
    \textbf{#1#2}\textit{\small #3}
}
\allowdisplaybreaks[4]
\numberwithin{equation}{section}

\definecolor{verde}{cmyk}{.83,.21,1,.08}
\definecolor{darkorchid}{rgb}{0.6, 0.2, 0.8}
\definecolor{darkgreen}{rgb}{0,.5,0}

\def\({\left(}
\def\){\right)}
\def\[{\left[}
\def\]{\right]}

\newcommand{\ii}{\mathrm{i}}

\newcommand{\dd}{\mathrm{d}}

\newcommand{\be}{\begin{equation}}
\newcommand{\ee}{\end{equation}}
\newcommand{\bea}{\begin{eqnarray}}
\newcommand{\eea}{\end{eqnarray}}
\newcommand{\del}{\partial}
\newcommand{\eqn}[1]{(\ref{#1})}

\newcommand{\la}{\label}

\newcommand{\R}{\mathbb R}

\newtheorem{proposition}{Proposition}[section]

\begin{document}
\title{Lie-Poisson gauge theories and $\kappa$-Minkowski electrodynamics }
\author[1]{V. G. Kupriyanov}
\author[2,3]{M. A. Kurkov}
\author[2,3]{P. Vitale}
\affil[ ]{}
\affil[1]{\textit{\footnotesize CMCC-Universidade Federal do ABC, 09210-580, Santo Andr\'e, SP, 
Brazil. }}
\affil[2]{\textit{\footnotesize Dipartimento di Fisica ``E. Pancini'', Universit\`a di Napoli Federico II, Complesso Universitario di Monte S. Angelo Edificio 6, via Cintia, 80126 Napoli, Italy.}}
\affil[3]{\textit{\footnotesize INFN-Sezione di Napoli, Complesso Universitario di Monte S. Angelo Edificio 6, via Cintia, 80126 Napoli, Italy.}}
\affil[ ]{}
\affil[ ]{\footnotesize e-mail: \texttt{vladislav.kupriyanov@gmail.com, max.kurkov@gmail.com, patrizia.vitale@na.infn.it}}
\maketitle
\begin{abstract}\noindent
%\blu{The Poisson gauge formalism naturally arises from the non-commutative gauge theory at the semi-classical approximation.}
{We consider  gauge theories on Poisson manifolds emerging as  semiclassical approximations of noncommutative spacetime with Lie algebra type noncommutativity.} We prove an important identity, which allows to obtain simple and manifestly gauge-covariant expressions for the Euler-Lagrange equations of motion, the Bianchi  and the Noether identities. We discuss the non-Lagrangian equations of motion, and apply our findings to the $\kappa$-Minkowski
case. We construct a family of exact solutions of the deformed Maxwell equations in the vacuum. In the classical limit, these solutions recover plane waves with  left-handed and  right-handed circular polarization, being  classical counterparts of photons. The deformed dispersion relation appears to be nontrivial.
\end{abstract}

\section{ Introduction}
{
Non-trivial Poisson structures over spacetime   may be regarded as semiclassical approximations of spacetime noncommutativity.  In view of that, the latter is conveniently  described   in terms of an associative  algebra of functions, algebraically dual to spacetime, made noncommutative by replacing the ordinary pointwise product with a noncommutative one (the star product). In such a context Poisson structures  have been widely investigated since the beginning of NC geometry. 

The definition  of a star product for
symplectic manifolds goes back to \cite{Bay, DeW-Lec, Fed}. The generalization to  Poisson manifolds  has been achieved by Kontsevich \cite{Konts}, whose star product reduces to the previous ones for nondegenerate structures.  A special family, which shall be the object of the present paper, is represented by  the so-called  Lie-Poisson structures, namely, linear Poisson brackets of Lie algebra type.  Because of their simple, but interesting mathematical structure, they  have been largely investigated  by the community. The first analysis to our knowledge is performed  in  \cite{gutt}. A far from complete list of subsequent works is represented by \cite{madore:1991bw, Hammou:2001cc, Gracia-Bondia:2001ynb, Lukierski:2005fc, Chryssomalakos:2007jr, Miao:2009jw, Meljanac:2014jsa}. 

Field and gauge theories on nontrivial Poisson manifolds, giving rise to noncommutative spacetimes which depart from   Moyal noncommutativity, have been  analysed within many different approaches \cite{Bimonte:1996fq, Arefeva:1999pkt, Jurco:2000fs, Amelino-Camelia:2001rtw, Amelino-Camelia:2002wis, Meusburger:2003hc, Banerjee:2007th, Kennedy:2012gk, Vitale:2012dz, Gere:2013uaa}; see for example \cite{Hersent:2022gry} for a recent review. A special mention deserves $\kappa$-Minkowski noncommutativity, also of Lie algebra type, which shall be considered in detail in  Sec. \ref{KappaES}.  

 There has been recently a renewed interest for Poisson gauge theories  within a novel approach which relies on strong homotopy algebras and symplectic embeddings \cite{Blumenhagen:2018kwq, Kupriyanov:2019cug, Kupriyanov:2020sgx,  Kupriyanov:2020axe, Kupriyanov:2021aet, Kupriyanov:2021cws, Kupriyanov:2022ohu,Abla:2022wfz}. Indeed,  while spacetime noncommutativity represents a natural solution to reconcile  general relativity and quantum mechanics in strong gravitational regimes \cite{Doplicher},  the state of the art of NC field and gauge theories is still unsatisfactory,  these being   generically affected by the so called  UV/IR  mixing \cite{mixing} or not reproducing  the standard commutative theories in the commutative limit, except for a few models with very special features.  

 The  present paper represents a followup  of previous work by the authors \cite{Kupriyanov:2020sgx,  Kupriyanov:2020axe, Kupriyanov:2021aet,  Kupriyanov:2022ohu}; it therefore shares the same motivations. We briefly review our arguments   along the lines of \cite{Kupriyanov:2020sgx}.

By Noncommutative Gauge Theory  (NCG)  it is generally intended a theory  described in terms of a noncommutative algebra $(\mathcal{A}, \star)$ representing spacetime, an $\mathcal{A}$-module,  $\mathbb{M}$, representing matter fields, a group of unitary automorphisms of $\mathbb{M}$  representing  gauge transformations.
The  dynamics of fields is  described by means of  a natural differential calculus based on  derivations  of the NC algebra. Moreover, the  gauge connection is  the standard noncommutative analog of the Koszul notion of connection \cite{segal, dbv1, dbv2, landimarmo, wesscalc, wallet1, wallet2} (for a physically inspired perspective   see the earlier  approach in    \cite{Madore:2000en}).

Therefore,   the first problem to face is the definition of an appropriate  differential calculus, namely, an algebra of $\star$-derivations  of $\mathcal{A}$ such that
\be
D_a (f\star g)= D_a f\star g+ f\star D_a g.
\ee
For constant noncommutativity, assuming $\Theta$  non-degenerate, the latter are given by star commutators 
\be 
D_a f= (\Theta^{-1})_{ab}[x^b, f]_\star\stackrel{\Theta\rightarrow 0}{\longrightarrow} \del_a f \label{der}
\ee
thus reproducing the correct commutative limit. 
For  coordinate dependent $\Theta(x)$ the situation is different. Lie algebra type star products, 
\be
[x^j, x^k]_\star= f^{jk}_l x^l \la{Lstar}
\ee
do admit a generalisation of \eqn{der} according to 
\be
D_j f \propto [x^j, f]_\star \label{dercom}
\ee
 but the limit  $\Theta\rightarrow 0$ is not directly related to the correct commutative result. In particular, in \cite{Vitale:2012dz} it is shown that, for $\mathfrak{su}(2)$ type noncommutativity, the differential calculus defined in terms of \eqn{dercom} becomes two-dimensional in the commutative limit. A related  approach  consists in using a  twisted differential calculus for those NC spacetimes which admit a twist formulation \cite{Vassilevich,chaichian1,chaichian2, aschieri, ALV08}. 

To summarise, for generic coordinate dependence of $\Theta$, where 
one needs to employ the general Kontsevich star product \cite{Konts}, 
\be \label{Kon}
f\star g=f\cdot g+\frac{i}{2}\,\Theta^{ab}(x)\,\partial_a f\partial_b g+\dots\,,
\ee
   ordinary derivations violate  the Leibniz rule,
\begin{equation*}
\partial_c(f\star g)=(\partial_c f)\star g+f\star(\partial_c g)+\frac{i}{2}\,\partial_c\Theta^{ab}(x)\,\partial_a f\partial_b g+\dots\,
\end{equation*}
whereas twisted or star derivations, although giving rise to a well defined differential calculus,  might not   reproduce the correct commutative limit (see for example \cite{Vitale:2012dz, Gere:2013uaa} where this point is analyzed in detail).  

In order to deal with this and other problems of NCG, in  the above mentioned papers \cite{Kupriyanov:2020sgx,  Kupriyanov:2020axe, Kupriyanov:2021aet,  Kupriyanov:2022ohu} 
 a different perspective is adopted:  the very definition of  gauge fields and gauge transformations is modified  in such a way that  not only the outcoming gauge theory  reproduce the correct commutative limit  but also, the gauge algebra close with respect to the underlying noncommutativity (namely, a homomorphism holds between the Lie algebra of gauge transformations and the noncommutative algebra of gauge parameters). 

To be precise, the whole construction is only performed in the semiclassical approximation where $\star$ commutators are replaced by Poisson brackets.  
  Strictly speaking this means that spacetime stays commutative, but it becomes a Poisson manifold with non-trivial Poisson bracket among position coordinates. Gauge parameters in turn, which are spacetime functions, inherit such a  non-trivial Poisson structure.  It is therefore  natural  to require that they close under  Poisson brackets. This implies a departure from the standard picture of NCG (and corresponding  Poisson gauge theory) outlined above, the latter being recovered only for  Moyal  noncommutativity, namely constant Poisson brackets. 
%%%%%%%%%%%%%%%%%%%%%%%%%%%%%%%

Before illustrating the new results of the present paper concerning $U(1)$  gauge theory on  a Poisson manifold (the spacetime) of   $\kappa$-Minkowski type,} we wish to add a few comments on the   motivations for considering   spacetimes  with a non-trivial Poisson structure and   argue that  such structures are conceivable and worth to be studied beyond their relationship with noncommutative geometry. One interesting instance is obtained by identifying  $\R^d$  with the dual, ${\mathfrak g}^*$, of some $d$-dimensional Lie algebra ${\mathfrak g}$.  ${\mathfrak g}^*$   carries the coadjoint  action of the group, whose orbits are homogenous spaces. There is a natural Poisson bracket on ${\mathfrak g}^*$, 
\be
\langle \{\ell^i, \ell^j\}, X_k\rangle= f^{ij}_k
\ee
of Lie algebra type, which is non-degenerate on the orbits [the Lie-Kirillov-Souriau-Konstant bracket (LKSK)]. Here  $\langle\;,\;\rangle$ indicates the natural pairing between ${\mathfrak g}$ and ${\mathfrak g} ^*$, $f^{ij}_k$ are the structure constants of the Lie algebra, $X_k\in {\mathfrak g}$ and $\ell^i$ their duals, identified with coordinate functions on ${\mathfrak g}^*$. This point of view is perfectly compatible with the standard approach to translation invariant dynamics, where the carrier space of the dynamics is the homogenous space of either Poincar\'e or Galilei group, respectively quotiented with respect to the Lorentz or Euclidean  subgroups. The induced LKSK is then trivial, being the dual of the translations commutator. 

Therefore, all situations where translation invariance is not the natural symmetry of spacetime, %e.g. curved spaces with some other symmetry at work, 
will have a natural Lie-Poisson bracket of the kind described above.

 Another interesting occurrence of nontrivial Poisson brackets on the configuration space is represented by dynamical systems defined on the manifold of some Lie group $G$. In that case, it is possible to endow the algebra of functions on $G$ with a Poisson bracket compatible with the group operations (in particular, the group multiplication has to be a Poisson map). This bracket, dubbed Poisson-Lie structure,\footnote{Poisson-Lie brackets are defined on the group manifold and should not  be confused with the Lie-Posson bracket of the dual Lie algebra.} is quadratic in the group coordinates and gives rise to the concept of Poisson-Lie group, the semi-classical counterpart of a quantum group \cite{Chari}.  Finally, it is worth mentioning that Poisson manifolds constitute an important family of target spaces for  topological field theories (see for example \cite{ikeda93, Schaller, Cattaneo:2001bp}).
 
 We will deal for most part of the paper with  linear Poisson brackets, although  more generally one could consider 
a smooth $d$-dimensional manifold~$\mathcal{M}$,  equipped with a  Poisson bracket of the form, 
 \begin{equation}
\{f,g\}=   {\Theta^{ab}(x)}\,\partial_a f\,\partial_b g\,  , \, \qquad f,g\in \mathcal{C}^{\infty}(\mathcal{M})
,
\la{pbr}
\end{equation}
where  $\Theta^{ab}(x)$ is a given Poisson bivector.\footnote{
Throughout the paper we use the notations of~\cite{Kupriyanov:2022ohu}.}
 
It is then natural to investigate the compatibility between the Poisson structure of the carrier manifold and the algebraic structure of the set of gauge transformations. In the following we shall consider a $U(1)$ gauge theory without matter fields, with  $A(x)$ representing the gauge one-form. 

The Poisson structure on the spacetime manifold induces a non-commutative algebra of infinitesimal gauge transformations for the gauge potential $A (x)$,
\begin{equation}
[\delta_f,\delta_g]A=\delta_{\{f,g\}}A \label{ga}
\end{equation}
whose bracket is dictated by the request that the gauge algebra be closed. 
{We also require that at the commutative limit  gauge transformations reproduce the standard $U(1)$ gauge transformations, 
\be
\lim_{\Theta\to 0}\delta_f A_j = \partial_j f. \la{ourcomlim}
\ee}
The matrix $\Theta $ shall be often called  the noncommutative  structure, although this is strictly speaking the Poisson tensor.

{We shall refer to field theoretical models with the gauge algebra satisfying Eqs. \eqn{ga}, \eqn{ourcomlim} as ~\emph{Poisson gauge theories}. As already noticed,   gauge theories on Poisson manifolds have been already studied in the literature in relation with  NCG and deformation quantization (see for example \cite{Bimonte:1996fq, Arefeva:1999pkt, Jurco:2000fs, Cattaneo:2001bp, Meusburger:2003hc, Kennedy:2012gk}). Towards  the end of the section we shall shortly comment on their relationship with our approach.}

 When restricting to 
Poisson structures 
of  Lie algebraic type, 
\be\label{liebr}
\Theta^{ab}(x) = f^{ab}_c x^c,
\ee
where the structure constants\footnote{$f^{ij}_k = \ii c^{ij}_k$, c.f. Eq.~\eqref{Lstar}.} $f^{ab}_c$ obey the Jacobi identity,
\be
f^{kl}_{i} f^{ja}_{l}  + f^{j l}_{i} f^{ak}_{l} + f^{al}_{i}f^{kj}_{l} = 0 \la{Jacobi}
\ee
the corresponding Poisson gauge theories will be referred to as    \emph{Lie-Poisson gauge theories}.

 In order for  gauge transformations to be compatible with the closure request \eqn{ga} and for the corresponding field strength to be gauge covariant, it has been shown~\cite{Kupriyanov:2021aet,Kupriyanov:2022ohu} that it is necessary to introduce a deformation of the gauge sector  in terms of two field-dependent matrices, $\gamma(A)$ and $\rho(A)$, which solve the master equations 
 \be
\left\{\begin{array}{l}
\gamma_{ i }^{ j } \partial^{ i }_A \gamma^{ k}_{ l} - \gamma^{ k}_{ i } \partial_A^{ i } \gamma^{ j }_{ l} 
  = \gamma^{ i }_{ l} f_{ i }^{ j  k} \\
\gamma^j_l \partial_A^l\rho_a^i \,+ \rho_a^l \partial_A^i\gamma_l^j   = 0  
\end{array}\right.
,
\la{firstsecond}
\ee
and obey the classical limits,
\be
\lim_{\Theta\to 0} \gamma = \mathbb{1},\qquad \lim_{\Theta\to 0} \rho = \mathbb{1}. \la{GRcomLim}
\ee
In parallel, a  gauge covariant derivative has been introduced in terms of the same objects. 
In Tab. \ref{tab1} we summarize the results and we refer to \cite{Kupriyanov:2021aet,Kupriyanov:2022ohu} for more details. We stress that, for a given Poisson structure,  the  matrices %\footnote{\blu{According to our notations, for any matrix  the upper index enumerates strings, whilst the lower one enumerates columns. }
  $\gamma(A)$ and $\rho(A)$ completely determine the fields and transformations of the related Poisson gauge model. \footnote{In general, the matrices $\gamma$ and $\rho$ may depend on $x$ explicitly. In the present work we limit the analysis to the $x$-independent case.}
\begin{table}
\begin{center}
\begin{tabular}{S|  |S} \toprule
    {Object}  & {Gauge-transformation rule}  \\ \midrule \midrule
  {$A_a(x)$}  & {$\delta_f A_a =\gamma^r_a(A)\,\partial_r f(x) +\{A_a ,f\}$} \\  \midrule
   {$\mathcal{F}_{ab}(x) = \rho_a^c(A)\,\rho_b^d(A)\big(\gamma_c^l(A)\,\partial_lA_d-\gamma_d^l(A)\,\partial_lA_c+\{A_c\,A_d\}\big)$}  & {$\delta_f {\cal F}_{ab}=\{{\cal F}_{ab},f\}$} \\  \midrule
    {${\cal D}_a\psi(x)=\rho_a^m(A)\, (\gamma_m^\ell(A) \del_\ell \psi+\{A_m,\psi\})$} & {$\delta_f\left({\cal D}_a\psi\right)=\{{\cal D}_a\psi,f\}$}    \\ \bottomrule
\end{tabular}
\caption{\label{tab1}{ Summary of the main objects and their transformation properties.  $\psi$ is an  arbitrary field which transforms in a covariant manner under gauge transformations: $\delta_f \psi =\{ \psi,f\} $. }}
\end{center}
\end{table}

Unless otherwise specified, the forthcoming analysis shall be restricted to the Lie-Poisson case \eqn{liebr}. To this, let us consider the gauge potential one-form $A= A_\mu dx^\mu$. %we  observe that the local basis of one-forms $\{dx^\mu\}$ inherits a Lie bracket  induced by the Poisson structure which reads  \cite{Koszul}
%\be
%[dx^i, dx^j]:= d\Theta^{ij}= f^{ij}_k dx^k.
%\ee
For the forthcoming discussion it is convenient to introduce the   following  notation%\footnote{This relation can be understood in terms of  the coadjoint representation of the Lie algebra on its dual, when the identification $\R^d\simeq \mathfrak{g}^*$ is made.} 
\be
\hat{A}^{s}_l = -\ii \,f_{l}^{is} A_i . \la{AhDef}
\ee
Two main remarks are in order before proceeding further. 
Firstly, the master equations \eqn{firstsecond} exhibit  \emph{universal} solutions in terms of matrix-valued functions~\cite{Kupriyanov:2021cws,Kupriyanov:2022ohu}: 
\be\la{universal}
\begin{array}{r@{}l}
\gamma(A) =& G(\hat{A}), \\
\rho^{-1}(A) =& \gamma - \ii\,\hat{A}. 
\end{array}
\ee
 which are valid for \emph{arbitrary} structure constants $ f^{ab}_c$.
In these formulae, 
\be
%\hat{A}^{s}_l &=& -\ii \,f_{l}^{is} A_i , \la{AhDef}\\
G(p) = \frac{\ii \, p}{2} + \frac{p}{2} \cot{\frac{p}{2}} = \sum_{n=0}^{\infty}  \frac{(\ii p)^n B_n^{-}}{n!},   \la{GDef}
\ee
and $B_n^{-}$, $n\in\mathbb{Z}_+$ stand for the Bernoulli numbers. The corresponding Poisson gauge theory will be addressed as the “universal" one.

Secondly, as shown in \cite{Kupriyanov:2022ohu}, any invertible field redefinition, 
\be
A\to\tilde A(A), \la{redef} 
\ee
such that,
\be
\lim_{\Theta\to 0}\tilde A(A) = A,
\ee
gives rise to new solutions of the master equations, 
\be
\tilde{\gamma}^{i}_j (\tilde{A}) = \Bigg(\gamma_k^i(A)\cdot \frac{\partial \tilde{A}_j}{\partial A_k}\Bigg)\Bigg|_{A = A(\tilde{A})}, \quad \tilde{\rho}_a^i(\tilde{A})  = \Bigg(\frac{\partial A_s}{\partial \tilde{A}_i}\cdot \rho_a^s(A)\Bigg)\Bigg|_{A = A(\tilde{A})}, \la{newGR}
\ee
which respect the commutative limits~\eqref{GRcomLim}.
In other words, for a given Poisson bivector $\Theta^{ab}$ one may construct infinitely many Poisson gauge theories. However, they are all equivalent: the field redefinitions map gauge orbits onto gauge orbits, thereby defining invertible  maps between these theories. {They   have been given the name of Seiberg-Witten maps in \cite{Kupriyanov:2022ohu}, because of some analogy with the famous result by Seiberg and Witten \cite{Seiberg:1999vs} in the framework of   fully noncommutative theory. (Notice however that a semiclassical analogue of the Seiberg-Witten result, closer to  the spirit of the original paper, is already discussed in\cite{Jurco:2000fs}; more details shall be given below).}  Interestingly,  all the Poisson gauge models which have been constructed so far within the present approach result to be connected with the universal one by means of Eqs.  \eqn{newGR}~\cite{Kupriyanov:2022ohu}.

{As mentioned before,  the case of nonconstant noncommutativity has been already addressed in the literature. As for example in the derivations based approach mentioned before,  the infinitesimal gauge transformation of the potential is usually assumed to be of  the form 
\be
\delta_f A = \alpha [f,  x]_\star+ [f, A]_\star
\ee
where  $\alpha $ is a suitable  constant, necessary to adjust the dimension. The semiclassical analogue is readily obtained by replacing star commutators with Poisson brackets.  It is easily seen that it satisfies  Eq. (\ref{ga}), but the commutative limit of the resulting gauge theory is not the standard one (see for example  the comment after Eq. \eqn{dercom} and \cite{Gere:2013uaa}). This expression, or, better to say, its semiclassical analogue,  coincides with the first  equation in Table \ref{tab1} only for  $\gamma_\mu^\nu= \delta_\mu^\nu$ which is a solution of  the first  master equation in  \eqn{firstsecond} only for $\Theta= const$,  while differing in the first term for all other cases. 
%Indeed, for constant $\Theta$ $\del_\mu f= \Theta_{\mu\nu}^{-1}  [x^\nu, f]_\star$. 
The same applies to the definition of the field strength $\mathcal{F}$ in the second row of the table: it agrees with previous definitions (e.g. those considered in \cite{Jurco:2000fs,Gere:2013uaa}) only if both $\gamma$ and $\rho$ are equal to the identity matrix, which is a solution for the second master equation only for $\Theta= const. $

The Poisson, or semiclassical analogue  of nonconstant noncommutativity has been addressed in \cite{Jurco:2000fs} in relation with the Seiberg-Witten map \cite{Seiberg:1999vs}. It is shown that a classical version of the latter may be established, such that, to the standard gauge theory with gauge transformation  $A\rightarrow A+ df$  on a classical manifold $\mathcal{M}$, it is possible to associate 
a one-parameter family of Poisson gauge theories with  gauge transformation
 \be
\delta_f A ^j_\rho= \{\tilde f,  x^j \}+ \{\tilde f, A^j_\rho \}.\, \la{dtrans}
\ee
In this formula  $A_\rho= A_\rho (A, f)$ and $\tilde f= \tilde f(A, f)$ are respectively the gauge potential  and gauge parameter  of the  family of Poisson gauge theories. Their results differ form the ones considered in this paper, as can be seen by comparing \eqn{dtrans} with the gauge transformation in Table \ref{tab1}.}

The paper is organised as follows.  In section \ref{firstres} we study the  covariance of the theory under gauge transformations. We first prove the explicit covariance of   the Bianchi identity previously obtained in \cite{Kupriyanov:2021aet}. Then  we present gauge-covariant expressions for the Euler-Lagrange equations of motion and the Noether identity, which exhibit correct commutative limits. Sec.~\ref{SecNLAgr} is dedicated to the non-Lagrangian gauge-covariant equations of motion for a generic Poisson tensor of  Lie-algebraic type. In Sec.~\ref{KappaES} we focus on the $\kappa$-Minkowski case,  introducing a one-parameter family of well-behaved non-Lagrangian equations of motion. We present the exact solutions of the deformed Maxwell equations in the vacuum, which recover plane waves with the left-handed and the right-handed polarization in the commutative limit. We conclude with a summary in section \ref{concl} and five appendices which contain technical aspects.

\section{Towards explicit covariance}\label{firstres}
The covariant field strength reported in Table \ref{tab1} has been shown to satisfy the following deformed Bianchi identity ~\cite{Kupriyanov:2021aet}   
\begin{equation}
{\cal D}_a{\cal F}_{bc}-{\cal F}_{ad}\,\mathcal{B}_b{}^{de}(A)\,{\cal F}_{ec} +\mbox{cycl}(abc)=0\,,  \la{Bianchi}
\end{equation}
with 
\begin{eqnarray}\label{F}
\mathcal{B}_b{}^{de}(A)&=&\left(\rho^{-1}(A)\right)_j^d\left(\partial^j_A\rho_b^m(A)-\partial^m_A\rho_b^j(A)\right)\left(\rho^{-1}(A)\right)_m^e\,.\label{Lambda}
\end{eqnarray}
%while $\mathcal{D}$ and $\mathcal{F}$ stand for the  gauge-covariant derivative and the gauge-covariant field strength respectively, as in Tab.\ref{tab1}. 
%see~\cite{Kupriyanov:2021aet} for details. being purely technical,
%is devoted to a very peculiar combination of the elements of $\rho$ and their derivatives, which appears in the Poisson gauge formalism, viz
Being an identity, Eq.~\eqref{Bianchi} must  be necessarily gauge covariant.  
However, the transformation properties of $\mathcal{B}_b{}^{de}(A)$ were not  explicitly  studied; in particular, we did not address the question as to whether or not  this quantity transforms in a covariant manner.  It is possible to prove that this is indeed the case. More precisely, the following Proposition holds
%In Sec.~{\bf a} of the Appendix we prove our main technical result, namely 
\begin{proposition}\label{prop1}
\begin{itemize} 
\item[]
\item[i)] In the universal Poisson gauge model \eqn{universal} it holds
\be
\mathcal{B}_c{}^{ab}(A) = f_c^{ab}. \la{mainres}
\ee 
\item[ii)]  Eq. \eqn{mainres} is valid for \emph{all} Poisson gauge theories, which are connected with the universal one through the Seiberg-Witten maps, i.e. the invertible field redefinitions~\eqref{redef} and the relations~\eqref{newGR}. 
\end{itemize}
\end{proposition}
The proof is explicitly given in Appendix \ref{propa}.
This Proposition  yields essential simplifications of the Lie-Poisson gauge theories.  In particular,  
%An immediate consequence of the relation~\eqref{mainres} is a 
we immediately get a manifestly gauge-covariant form of the deformed Bianchi identity,
\be
{\cal D}_a{\cal F}_{bc} -{\cal F}_{ad}\,f_b{}^{de} \,{\cal F}_{ec} +\mbox{cycl}(abc)=0.
\ee
Moreover, combining {the results of the Proposition 5.1 in~\cite{Kupriyanov:2021aet} which provides the commutation relation of  gauge covariant derivatives,
\be
{[\mathcal{D}_a,\mathcal{D}_b] = \left\{\mathcal{F}_{ab},\cdot\right\} + \mathcal{F}_{ad}\,\mathcal{B}^{de}_b(A)\,\mathcal{D}_e 
- \mathcal{F}_{bd}\,\mathcal{B}^{de}_a(A)\,\mathcal{D}_e\,} \la{comDp}
\ee
 with Proposition \ref{prop1}, we arrive at the following simple formula for the commutator of the covariant derivatives:
\be
[\mathcal{D}_a,\mathcal{D}_b] = \left\{\mathcal{F}_{ab},\cdot\right\} + \mathcal{F}_{ad}\,f^{de}_b\,\mathcal{D}_e 
- \mathcal{F}_{bd}\,f^{de}_a\,\mathcal{D}_e. \la{comD}
\ee

Therefore, we shall assume in the rest of the paper  that the  Poisson gauge model under analysis is either the universal one, or it is connected with the universal model via a Seiberg-Witten map, i.e. the identity~\eqref{mainres} applies.

\subsection{ Lagrangian dynamics. \la{SecLAgr}}
Any Poisson gauge theory exhibits a Lagrangian formulation. As we shall see below, some of these formulations have a correct commutative limit, while others do not.
\subsubsection{Admissible Lagrangian models}
The classical action,
\be
S_g[A] = \int_{\mathcal{M}} \dd x \,\Big(-\frac{1}{4}\mathcal{F}_{ab}\,\mathcal{F}_{cd}\, \eta^{ca}\,\eta^{bd} \Big), \qquad \eta = \mathrm{diag}(\,+1,-1,\dots,\,-1), \la{gaction}
\ee
$\mathcal{M}= \R^d$, is gauge-invariant 
if and only if the Poisson tensor obeys  the \emph{compatibility condition}~\cite{Kupriyanov:2020axe},
\be
\partial_{a} \Theta^{ab}=0 . \la{LtyRel}
\ee
For $\Theta$ of the Lie-algebra type this condition is equivalent to the following restriction on the structure constants,
\be
f^{ab}_a = 0. \la{LtyRel1}
\ee
 We shall call these Lagrangian formulations the “\emph{admissible}" ones. 
 All the special Lie algebras satisfy \eqn{LtyRel1}, hence they  yield admissible Lagrangian formulations of the corresponding Lie-Poisson gauge theories. Below we present manifestly gauge-covariant expressions for the equations of motion and for the Noether identity. 

\begin{proposition}\label{prop2}
For  admissible Lagrangian models the Euler-Lagrange equations of motion,
\be
\mathcal{E}_{EL}^{a} = 0, \qquad  \mathcal{E}^a_{EL} := \frac{\delta S_g}{\delta A_a},
\ee
can be rewritten in a manifestly gauge-covariant way,
\be
\mathcal{E}_{C}^c = 0, \qquad \mathcal{E}^a_{C} :=
 \mathcal{D}_c \mathcal{F}^{ca} + \frac{1}{2}\, \mathcal{F}_{de}\,f_b^{de}\, \mathcal{F}^{ab} -\mathcal{F}_{de} \, f_b^{ae}\,\mathcal{F}^{db}. 
\ee
\end{proposition}
\emph{Proof.} 
Using the results of~\cite{Kupriyanov:2021aet}, accompanied by Proposition \ref{prop1}, one can easily see that,
\be
\mathcal{E}_{EL}^{a} = \rho^a_b \mathcal{E}_{C}^{b}.  \la{CE0}
\ee
The matrix $\rho$ is non-degenerate, therefore $\mathcal{E}_{EL}^{a}=0$ iff $\mathcal{E}_{C}^{b}=0$. Q.E.D. \\ \\
\noindent\emph{Remark}. In the three-dimensional case with the $su(2)$-non-commutativity, the combination
\be
\frac{1}{2}\, \mathcal{F}_{de}\,f_b^{de}\, \mathcal{F}^{ab} -\mathcal{F}_{de} \, f_b^{ae}\,\mathcal{F}^{db}
\ee
vanishes, and one arrives to the so-called “natural" equations of motion $ \mathcal{D}_c \mathcal{F}^{ca} =0$, see~\cite{Kupriyanov:2022ohu} for details.
However one can check by  direct calculation, that the addition of a fourth commuting coordinate (e.g. time) breaks this property. A non-triviality due to the presence of commutative coordinates in the Poisson gauge formalism was discovered in~\cite{Kurkov:2021kxa}.

\begin{proposition}\label{prop3}
If an admissible Lagrangian formulation of the Poisson gauge theory is available, the Noether identity~\cite{Kupriyanov:2022ohu},
\be
\partial_l (\mathcal{E}_{EL}^a \gamma_a^l(A))  + \{A_a, \mathcal{E}_{EL}^a\} =0, \la{NTi}
\ee 
can be rewritten in a manifestly gauge-covariant form,
\be
\mathcal{D}_a \,\mathcal{E}^a_{C} -\frac{1}{2} \, f_{b}^{rt}\, \mathcal{F}_{rt} \,\mathcal{E}^b_{C} = 0. \la{CEconnect}
\ee
\end{proposition}
\noindent\emph{Proof.} 
Substituting the relation~\eqref{CE0} in the left-hand side of the Noether identity~\eqref{NTi}, we obtain, 
\be
\mathbf{l.h.s.} =\partial_l (\mathcal{E}_{EL}^a \gamma_a^l(A))  + \{A_a, \mathcal{E}_{EL}^a\} 
= \mathcal{D}_a\mathcal{E}^a_C 
+ \big(\partial_l (\gamma_a^l \rho_b^a) + \left\{A_a, \rho_b^a\right\}\big)\,\mathcal{E}^b_C.
\ee
where, on using the  definition of $\mathcal{D}_a$ as in Table \ref{tab1},
\be
\mathcal{D}_b\mathcal{E}^a_C = \rho_b^d\,\big(\gamma_d^l \partial_l \mathcal{E}^a_C  + \left\{A_d,\mathcal{E}^a_C \right\}\big).
\ee
In App. \ref{appbc} we prove that,
\bea
 \left\{A_a, \rho_b^a\right\} &=& -\frac{1}{2}\, f_b^{rt}\,\big(\rho_r^q\, \rho_t^l \,\left\{A_q,A_l\right\}\big),  \la{PCstep1}\\
\partial_l (\gamma_a^l \rho_b^a) &=&  -\frac{1}{2}\, f_b^{rt}\, \rho_r^q\,\rho_t^l\,(\gamma_q^p\,\partial_p A_l - \gamma_l^p\,\partial_p A_q). \la{PCstep2}
\eea
Therefore, by using the definition of $\mathcal{F}$, we  arrive at, 
\be
\mathbf{l.h.s.} =  \mathcal{D}_a\mathcal{E}^a_C   -\frac{1}{2}\, f_b^{rt}\,
\underbrace{\rho_r^q\, \rho_t^l\,\big( \gamma_q^p\,\partial_p A_l - \gamma_l^p\,\partial_p A_q + \left\{A_q,A_l\right\}\big)}_{\mathcal{F}_{rt}}
\,\mathcal{E}^b = \mathcal{D}_a \,\mathcal{E}^a_{C} -\frac{1}{2} \, f_{b}^{rt}\, \mathcal{F}_{rt} \,\mathcal{E}^b_{C} ,
\ee
which is nothing but the left-hand side of Eq.~\eqref{CEconnect}. Q.E.D.

\subsubsection{Non-admissible Lagrangian models}
%In the previous section we assumed that the non-commutativity obeys the compatibility condition~\eqref{LtyRel}, i.e. the the “good" %Lagrangian formulation of the corresponding Poisson gauge theory was available.  
In some relevant cases, such as  the $\kappa$-Minkowski non-commutativity, 
\be
\Theta^{ab}(x) = \lambda\,\big(\delta_0^a\, x^b - \delta_0^b\,x^a \big), \la{ThetaKappa}
\ee
the compatibility condition~\eqref{LtyRel} is not satisfied. Therefore, the corresponding action is not gauge invariant. 
One can still define a gauge-invariant classical action, modifying the integration measure by a function $\mu$, as follows 
\be
S_g[A] = \int_{\mathcal{M}} \dd x \,\mu(x)\,\Big(-\frac{1}{4}\mathcal{F}_{ab}\,\mathcal{F}_{cd}\, \eta^{ca}\,\eta^{bd} \Big) 
\ee
 with the  measure  function  obeying  the  condition
\be
\partial_a \,\big(\mu(x)\, \Theta^{ab}(x) \big) = 0  \la{muEq}
\ee
see~\cite{Kupriyanov:2020axe} for details.
For  $\kappa$-Minkowski non-commutativity the general solution of Eq.~\eqref{muEq} reads\footnote{{One has to set $a^0 = \lambda/2$, $a^1=0$, $a^2=0$,  ..., $a^{d-1}=0$ in Eq.(4.15) of the quoted reference.}}~\cite{Kupriyanov:2020axe},
\be\label{muhigher}
\mu(x,\lambda)  = \frac{1}{( x^1 )^{d-1}}\cdot F_{\lambda}\left(
\frac{ x^2 }{ x^1 }, \frac{ x^3 }{ x^1 },..., \frac{ x^{d-1} }{ x^1 }
\right),
\ee
where $F_{\lambda}(w^2,...,w^{d-1})$ is an arbitrary, sufficiently regular, function of $d-2$ variables. We added the subscript $\lambda$ %in the notation of this function, 
in order to emphasise that $F$  may also depend on the deformation parameter $\lambda$ in an arbitrary manner.

However, by a simple scaling argument, presented in App.~\ref{appd}, it can be checked that this solution is not compatible with the commutative limit\footnote{At $d\neq4$ the measure $\mu$ can be absorbed by a conformally-flat metric tensor $g_{ab}(x) = \mu^{\frac{2}{d-4}}\eta_{ab}$.  In particular, at $d=2$ a flat commutative limit of the action is possible, even though the condition~\eqref{mucommlim} is not satisfied, see~\cite{Kupriyanov:2020axe} for details. In this case the coordinate dependent metric tensor $g_{ab} $ is a flat metric, written in curvilinear coordinates.  Nevertheless,  at $d=4$ the Maxwell action is conformally invariant, therefore  such a way out is not applicable, and the condition~\eqref{mucommlim} is strictly necessary. }
\be
\lim_{\Theta\to 0} \mu(x) = 1. \la{mucommlim}
\ee
Thus, since the measure $\mu$ enters in the Euler-Lagrange equations of motion, the corresponding Lagrangian dynamics does not reproduce the commutative limit correctly. We shall refer to these models as non-admissible. Therefore we are strongly motivated to look for gauge-covariant equations of motion, which lie beyond the Euler-Lagrange formalism.
%%%%%%
%QQQQQ
%%%%%%

\subsection{ Non-Lagrangian dynamics \la{SecNLAgr}}
By contracting the covariant derivative $\mathcal{D}_a$ with the covariant field-strength $\mathcal{F}_{ab}$ and the structure constants 
$f^{ab}_c$, one can obtain various e.o.m., which reproduce the standard dynamical sector of Maxwell equations
\be
\mathcal{E}^{a}_{\mathrm{Maxwell}} =0,\qquad \mathcal{E}^a_{\mathrm{Maxwell}} := \partial_b \,F^{ba} , \qquad F_{ab} = \partial_aA_b-\partial_b A_a,   \la{Maxwell}
\ee
 at the commutative limit. As we shall see below, not all  these opportunities are equally good.
\subsubsection{Natural equations of motion}
The simplest possibility for   gauge-covariant e.o.m. is,
\be
\mathcal{E}_N^a =  0,\qquad \mathcal{E}_N^a := \mathcal{D}_b\mathcal{F}^{ba} . \la{natural}
\ee
These are the so called “natural" equations of motion. Being manifestly gauge-covariant, these e.o.m. reduce to the  Maxwell equations~\eqref{Maxwell} in the commutative limit.

However, it is possible to show that there is no one-to-one correspondence between the solutions of classical Maxwell equations and the solutions of ~\eqref{natural}. Indeed, on contracting the natural e.o.m.~\eqref{natural} with the covariant derivative $\mathcal{D}_a$, we obtain the following equality, 
which has to be satisfied on shell\footnote{i.e. when the field $\mathcal{F}$ satisfies the e.o.m.},
\bea
[\mathcal{D}_a,\mathcal{D}_b]\mathcal{F}^{ab}  = 0. \la{bug0}
\eea
%%%%%
%QQQQQ
%%%%%
Hence, the identity~\eqref{comD} yields
\be
 \mathcal{F}_{ad}\,f^{de}_b\,\mathcal{D}_e \,\mathcal{F}^{ab} = 0. \la{bug1}
\ee
In order to perform the commutative limit,  it is convenient to rescale  the structure constants by the deformation parameter $\lambda$ according to 
\be
f_c^{ab} = \lambda \hat{f}_c^{ab},
\ee 
so that  $\hat{f}_c^{ab}$ be pure numbers.
Let us assume that the solutions of the deformed e.o.m.~\eqref{natural} tend to  solutions  of the standard Maxwell e.o.m~\eqref{Maxwell}
in  the  limit $\lambda\to 0$, namely that,  \emph{on-shell},
\be\label{onshell}
\lim_{\lambda\to 0} \mathcal{F}_{ab}  = F_{ab}. 
\ee
By construction  such a limit  is automatically verified off-shell. However, a priori nothing guarantees that the deformed solutions are analytic functions of $\lambda$.  In other words, the assumption \eqn{onshell}  is not an obvious property at all.
On performing the   limit  $\lambda\to 0$ in Eq.~\eqref{bug1}, we arrive at the \emph{nonlinear} constraint,
\be
 F_{ad}\,\hat{f}^{de}_b\,\partial_e F^{ab} = 0, \la{bug}
\ee
which, in general, is \emph{not} automatically satisfied by  the solutions of  the classical Maxwell equations~\eqref{Maxwell}.  Therefore, any solution of the commutative e.o.m., which does not obey this additional constraint, does not have a deformed counterpart solving  the natural e.o.m.~\eqref{natural}.

\subsubsection{Covariant equations of motion} \label{covsec}
Appropriate  equations of motion, which do not suffer the problem mentioned above, can be formulated. They are indeed   the covariant equations of motion introduced in the Euler-Lagrange context, namely
\be{
\mathcal{E}_{C}^a = 0, \qquad \mathcal{E}^a_{C} :=
 \mathcal{D}_c \mathcal{F}^{ca} + \frac{1}{2}\, \mathcal{F}_{de}\,f_b^{de}\, \mathcal{F}^{ab} -\mathcal{F}_{de} \, f_b^{ae}\,\mathcal{F}^{db}. \la{CovEOM} }
\ee
These equations remain valid even when  the compatibility condition~\eqref{LtyRel} is not fulfilled (e.g. in the $\kappa$-Minkowski case). In such cases  they  can \emph{not} be obtained from an action principle, but have to be stated independently.
These equations enjoy the following  important property.
\begin{proposition}\label{propD}
The left-hand-side of the covariant e.o.m. \eqn{CovEOM} satisfies 
\be
\mathcal{D}_a \,\mathcal{E}^a_{C} =\frac{1}{2} \, f_{c}^{ab}\, \mathcal{F}_{ab} \,\mathcal{E}^c_{C} \la{usfulidentity}
\ee
for \emph{all} non-commutativity of  Lie-algebra type.
\end{proposition}

The proof is presented in App. \ref{appe}.  As we have seen with Prop. \ref{prop3}  this formula is nothing but the Noether identity, when an admissible   Lagrangian formulation of the Poisson gauge theory is available. 
We emphasise, however, that Eq.~\eqref{usfulidentity} is valid always, no matter whether the compatibility relation~\eqref{LtyRel} takes place or not. 
From now on, by somewhat stretching the terminology, we shall call Eq.~\eqref{usfulidentity} the “Noether identity" in the non-Lagrangian cases as well.

As we announced above, thanks to the latter, no extra constraints in the classical limit arise.  
Indeed, by contracting the covariant e.o.m.~\eqref{CovEOM} with $\mathcal{D}_c$,  %dividing the outcoming equality by $\lambda$ 
and passing to the limit $\lambda\to 0$, we get the relation
 \be
\hat{ f}_{c}^{ab}\, F_{ab} \,\partial_d F^{dc} = 0,
 \ee 
which is automatically satisfied for \emph{all} solutions of  Maxwell equations~\eqref{Maxwell}. In contrast to the natural e.o.m~\eqref{natural}, the covariant ones~\eqref{CovEOM}  do not yield any special constraints.

{Let us clarify the importance of the Noether identity. In the commutative case not all of the dynamical Maxwell equations are independent.
The constraint is given by the Noether identity, 
\be
\partial_a \mathcal{E}_{\mathrm{Maxwell}}^a = 0. \la{MaxNoeth}
\ee
 On one hand, the natural equations of motion~\eqref{natural} do not yield a gauge-covariant Noether identity  respecting the commutative limit. From this point of view, \eqref{natural}, accompanied by the requirement of the correct commutative limits of their solutions, are overdetermined. On the other hand, the covariant e.o.m. do exhibit a  gauge-covariant Noether identity, which reduces to the constraint~\eqref{MaxNoeth} in the commutative limit. Therefore, the covariant e.o.m. are not overdetermined in the above mentioned sense.
 \\

%{\noindent\emph{{\bf \emph{Open question.}} %The natural e.o.m. exhibit problems at the linear order in $\lambda$. 
% It would be nice to demonstrate that for any solution of the Maxwell's equations~\eqref{Maxwell}, one can build its deformed counterpart, which solves the covariant e.o.m.~\eqref{CovEOM}. This is definitely a complicated task.  We hope, the Noether's identity~\eqref{usfulidentity} will help us again.  }}

\subsection{Generalised equations of motion}
Summarising the previous analysis, we see that the covariant e.o.m are, generally speaking, non-Lagrangian e.o.m, with the following   properties:
\begin{itemize}
\item{{\bf Correct commutative limit:} the e.o.m reproduce the first pair~\eqref{Maxwell} of Maxwell equations at the commutative limit. }
\item{{\bf Consistency with the Lagrangian formalism:} the e.o.m  can be obtained from the classical action~\eqref{gaction}, when an admissible Lagrangian formulation of the corresponding Poisson gauge theory is available.}
\item{{\bf Minimality:} The only term which contains the covariant derivative is $\mathcal{D}_c\mathcal{F}^{ca}$. The derivative-free terms are  quadratic in $\mathcal{F}$. }
\item{{\bf Noether identity:} Prop. \ref{propD} holds. This property allows to avoid the pathologies discussed at the beginning of this section in the context of the “natural" e.o.m.}
\end{itemize}
In what follows we rise the  question as to whether    it possible to generalise the covariant e.o.m, preserving the  properties listed above.  In other words, we wonder whether there are other non-Lagrangian e.o.m, which are as good as the covariant ones.} 

The following two-parameter family of  generalised e.o.m. satisfies the first three requirements,
\be
\mathcal{E}^a_G=0,\qquad \mathcal{E}^a_G =  \mathcal{E}^a_C + \alpha\, f_{b}^{ba}\, \mathcal{F}_{ed}\,\mathcal{F}^{ed} 
+ \beta\,f_b^{be}\,\mathcal{F}_{ed}\,\mathcal{F}^{da},\qquad \alpha,\beta \in \mathbb{R}.  \la{genEOM}
\ee
 When the compatibility condition~\eqref{LtyRel1} is fulfilled, the new terms vanish identically and we end up with the covariant e.o.m. The corresponding Noether identity may also contain new terms,
\be
\mathcal{D}_a \,\mathcal{E}^a_{G} =\left(\frac{1}{2} \, f_{c}^{ab}\, \mathcal{F}_{ab} + \omega\,f_a^{ab}\,\mathcal{F}_{bc}\right) \,\mathcal{E}^c_{G},\qquad \omega\in\mathbb{R}, \la{usfulidentityG}
\ee
which, of course,  vanish, when an admissible Lagrangian formulation is available. In the next section, analysing the $\kappa$-Minkowski non-commutativity, we shall see that these generalisations are indeed possible. Moreover, there exist  special choices of the constants 
$\alpha$, $\beta$ and $\omega$, which simplify the analysis of the corresponding dynamics.

\section{Applications to the $\kappa$-Minkowski non-commutativity} \la{KappaES}
%\subsection*{a. Simple $\kappa$-Minkowski Poisson gauge model.}
Throughout this section we consider the $\kappa$-Minkowski non-commutativity~\eqref{ThetaKappa} at $d=4$, which is probably the most well studied example of non-commutative space with non-trivial Poisson structure~\cite{lukierski}-\cite{kappa2}. The corresponding structure constants read,
\be
f^{ab}_c = \lambda\,\big(\delta_0^a\, \delta^b_c - \delta_0^b\,\delta^a_c \big). \la{fKappa}
\ee
One can easily check that the  matrices,
\be
\gamma(A)  =  \left(
\begin{array}{cccc}
1 &-\lambda A_1 &-\lambda A_2 & -\lambda A_3 \\
0 &1 &0 &0 \\
0 &0 &1 &0 \\
0 &0 &0 &1 
\end{array}
\right), 
\qquad 
\rho(A)  =  \left(
\begin{array}{cccc}
1 &0 &0 & 0 \\
0 & e^{\lambda A_0} &0 &0 \\
0 &0 &e^{\lambda A_0} &0 \\
0 &0 &0 &e^{\lambda A_0} 
\end{array}
\right), \la{newRG}
\ee
solve the master equations~\eqref{firstsecond}\footnote{{Let us note that the corresponding symplectic embedding $\gamma(p)$ was already known in a different context \cite{%Lukierski,
Kowalski-Glikman:2002eyl, Kowalski-Glikman:2002iba}.%, however tha expression for the matrix $\rho(A)$ needed for the construction of the consistent dynamics of the gauge field is absolutely new.
}}. 

For this kind of noncommutativity, Poisson gauge models,  including the universal one,  were already obtained 
 in previous publications ~\cite{Kupriyanov:2020axe,Kupriyanov:2021aet,Kupriyanov:2022ohu}. They  were shown to be related with each other through Seiberg-Witten maps~\cite{Kupriyanov:2022ohu}. 
 In  the following we shall present a new model along the lines described above.  In  Appendix \ref{appf} we prove   that it is equivalent via  Seiberg-Witten maps to the previous ones  as well.

\subsection{Generalised equations of motion}
One can check by  direct substitution that the generalised e.o.m.~\eqref{genEOM} are compatible with the (generalised) Noether identity~\eqref{usfulidentityG} iff,
\be
\beta = 4\,\alpha,  \quad \omega =  -4\,\alpha.
\ee
We have a one-parameter family of equally acceptable generalised e.o.m., whose left-hand sides, 
\be
\mathcal{E}^a_G =  %\mathcal{E}^a_C 
 \mathcal{D}_c \mathcal{F}^{ca} + \frac{1}{2}\, \mathcal{F}_{de}\,f_b^{de}\, \mathcal{F}^{ab} -\mathcal{F}_{de} \, f_b^{ae}\,\mathcal{F}^{db} + \alpha\,\big( f_{b}^{ba}\, \mathcal{F}_{ed}\,\mathcal{F}^{ed} 
+ 4\,f_b^{be}\,\mathcal{F}_{ed}\,\mathcal{F}^{da}\big),  \qquad \alpha \in \mathbb{R}, \la{kGeom}
\ee
obey the (generalised) Noether identity,
\be
\mathcal{D}_a \,\mathcal{E}^a_{G} =\frac{1}{2}\cdot\Big(1+12\,\alpha\Big) \, f_{c}^{ab}\, \mathcal{F}_{ab}   \,\mathcal{E}^c_{G}.
\ee
In order to obtain this formula, we used the equality,
\be
f_a^{ab}\,\mathcal{F}_{bc} = -\frac{3}{2} \, f_{c}^{ab}\, \mathcal{F}_{ab} ,
\ee
which is a peculiar feature of the $\kappa$-Minkowski non-commutativity.
Notice that, at $\alpha = -1/12$,  the Noether identity takes more elegant form, 
\be
\mathcal{D}_a \,\mathcal{E}^a_{G} = 0 . 
\ee
As we shall see soon, this choice of $\alpha$, indeed, simplifies the analysis of the  nonlinear equations of motion. 

\subsection{ $\kappa$-Minkowski plane waves}
From a particle physics perspective, the most interesting solutions of  Maxwell equations in the vacuum are the plane waves with the left-handed and the right-handed circular polarizations. In Quantum Electrodynamics these waves correspond to photons with the left-handed and the right-handed helicities respectively.  Poisson gauge models with the $\kappa$-Minkowski non-commutativity are rotationally invariant\footnote{The rotational invariance follows from the explicit expressions of the Poisson tensor  (\ref{ThetaKappa}), the  matrices $\gamma(A)$ and $\rho(A)$ given by (\ref{newRG}) and the corresponding equations of motion. Moreover, the theory is related to the abelian twist which respects  rotational invariance \cite{ML1%ML2
}.}, therefore, without loss of generality, we consider plane waves, travelling along the axis $x^3$. 
First, to set the notations, we present the relevant formulae of the commutative electrodynamics. Then we  focus on the non-commutative case. 
\subsubsection{Circularly polarized plane waves in the commutative case.}
In the Coulomb gauge, the four-potential describing these  waves  is given by
\be
A^{\pm}_{1\,{\bf cl}}  =  \varepsilon \, \sin{(k^0\,x^0 - k^3 \,x^3)},\qquad A^{\pm}_{2\,{\bf cl}} = \mp \,\varepsilon \, \cos{(k^0\,x^0 - k^3 \,x^3)}  \qquad A_{0\,{\bf cl}}^{\pm} = 0 =A^{\pm}_{3\,{\bf cl}},  \la{Acirc}
\ee 
where $\varepsilon$ stands for its amplitude, $k$ denotes  the wave vector, and the signs plus and minus specify the polarisations, see below for details. Hereafter the subscript  “$\bf cl$" indicates that we are dealing with the “classical" solution, which refers to the undeformed Maxwell electrodynamics.  The gauge potential~\eqref{Acirc} solves the Maxwell e.o.m. iff the wave vector satisfies the dispersion relation,
\be
k_0^2 -\vec{k}^2=0. \la{disprel}
\ee

The corresponding electric and magnetic fields, $\vec{E}^{\pm}_{\bf cl}, % = (E_{1 {\bf cl}}^{\pm }, E_{2 {\bf cl}}^{\pm }, E_{3 {\bf cl}}^{\pm })$ and 
\vec{B}^{\pm}_{\bf cl}% = (B_{1 {\bf cl}}^{\pm }, B_{2 {\bf cl}}^{\pm}, B_{3 {\bf cl}}^{\pm})
$ read,
\bea
\vec{E}^{\pm}_{\bf cl} &=& -\partial_0\, \vec{A} \,= \varepsilon\, k^0\cdot\big(+ \cos{(k^0\,x^0 - k^3 \,x^3)}, \pm\, \sin{(k^0\,x^0 - k^3 \,x^3)},0 \big), \nonumber\\
\vec{B}^{\pm}_{\bf cl} &=& \vec{\nabla}\times \vec{A} =  \varepsilon\, k^3\cdot\big( \mp\, \sin{(k^0\,x^0 - k^3 \,x^3)}, +\cos{(k^0\,x^0 - k^3 \,x^3)},0 \big), \la{Cwaves}
\eea
%where, by definition, 
%\bea
%\vec{A}&:=& (A^1,A^2,A^3) = -(A_1,A_2,A_3), %\nonumber \\
%%\boldsymbol{\varepsilon} &:=& k_0\cdot \varepsilon, 
%\eea
%and $\vec{\nabla}$ stands for the three-dimensional gradient.
If $\mathrm{sgn}(k^0) = \mathrm{sgn}(k^3) = 1$, then the fields $(\vec{E}_+, \vec{B}_+)$ and $(\vec{E}_-, \vec{B}_-)$ describe the right-handed and the left-handed  circularly polarized plane waves respectively.

\subsubsection{Non-commutative deformations.}
%Consider a special realisation of the $\kappa$-Minkowski Poisson gauge model, which corresponds to the special choice $\alpha=-1/12$, mentioned above. One can check by direct substitution, that the gauge potentials,
By substituting the Ansatz,
\be
{{A^{\pm}_{1}  =  \varepsilon \, \sin{(k^0\,x^0 - k^3 \,x^3)},\qquad A^{\pm}_{2} = \mp \,\varepsilon \, \cos{(k^0\,x^0 - k^3 \,x^3)},  \qquad A_{0}^{\pm} = 0 =A^{\pm}_{3},} } \la{AcircNC}
\ee 
in the non-commutative equations of motion,
\be
{\mathcal{E}^a_G = \mathcal{D}_c \mathcal{F}^{ca} + \frac{1}{2}\, \mathcal{F}_{de}\,f_b^{de}\, \mathcal{F}^{ab} -\mathcal{F}_{de} \, f_b^{ae}\,\mathcal{F}^{db} +\alpha\,\big( f_{b}^{ba}\, \mathcal{F}_{ed}\,\mathcal{F}^{ed} 
+ 4\,f_b^{be}\,\mathcal{F}_{ed}\,\mathcal{F}^{da}\big)  = 0,}
\ee 
we obtain,
\bea
\mathcal{E}_G^0 &=&-\lambda\, \big[  \left( 2\,{\lambda}^{2} \left( 1+3\,\alpha \right) 
{\varepsilon}^{2}+6\,\alpha-1 \right) {k_{{0}}}^{2}+ \left( 6
\alpha+2 \right) {k_{{3}}}^{2} \big] {\varepsilon}^{2}, \nonumber\\
 \mathcal{E}^{1,2}_G &=& A_{1,2}\cdot \big[\big(1-2\lambda^2(6\alpha +1)\varepsilon^2\big)k_0^2 - k_3^2\big], \nonumber\\
  \mathcal{E}^{3}_G &=&- \left( 12\alpha+1 \right) \lambda\,{\varepsilon}^{2}k^{{0}}
k^{{3}}.
\eea
At $\alpha = -1/12$ the expressions~\eqref{AcircNC}, indeed, solve the e.o.m., 
iff the \emph{deformed} dispersion relation, 
\be
{\big(1-\lambda^2\,\varepsilon^2\big) \,k_0^2 -\big(\vec{k}\big)^2 = 0,} \la{defdisprel}
\ee 
is fulfilled.  
The structure of the deformed gauge potential is identical to the commutative one~\eqref{Acirc}.
For other choices of $\alpha$ the simple Ansatz \eqn{AcircNC}  does not work, and one has to try  more involved ones, what goes beyond the scope of the present work. {Note that the deformation of the dispersion relation in the scalar field theory on the $\kappa$-Minkowski space constructed from the abelian twist was obtained in \cite{Pachol,Pachol1}, however in that approach no correction was produced  for the vector fields.}

The corresponding deformed electric  and magnetic fields, 
\be
 \vec{E}  := (\mathcal{F}_{01},\mathcal{F}_{02},\mathcal{F}_{03}), \qquad \vec{B} := (\mathcal{F}_{32},\mathcal{F}_{13} ,\mathcal{F}_{21}),
 \ee
read,
{\bea
\vec{E}^{\pm} &=& \varepsilon\, k^0\cdot\big(+ \cos{(k^0\,x^0 - k^3 \,x^3)}, \pm\, \sin{(k^0\,x^0 - k^3 \,x^3)},0 \big), \nonumber\\
\vec{B}^{\pm} &=&  \big( \mp\, \varepsilon\, k^3\, \sin{(k^0\,x^0 - k^3 \,x^3)}, + \varepsilon\, k^3\,\cos{(k^0\,x^0 - k^3 \,x^3)}, \pm \lambda\,\varepsilon^2\,k_0 \big). \la{NCwaves}
\eea}
In the commutative limit all the deformed quantities tend to their classical counterparts,
\be
\lim_{\lambda\to 0} A^{\pm} = A^{\pm}_{\bf cl},\qquad  \lim_{\lambda\to 0} \vec{E}^{\pm} = \vec{E}^{\pm}_{\bf cl},\qquad \lim_{\lambda\to 0} \vec{B}^{\pm} = \vec{B}^{\pm}_{\bf cl}.
\ee
Though the expressions~\eqref{NCwaves} for the deformed electromagnetic field components
are very similar to the ones for their commutative analogs~\eqref{Cwaves},  there are two important differences.
\begin{itemize}
{\item{The dispersion relation is the deformed one, given by Eq.~\eqref{defdisprel}. }
\item{The deformed magnetic field $\vec{B}^{\pm}$ has non-zero longitudinal components $B_z^{\pm} =\pm \lambda\,\varepsilon^2\,k_0  $, therefore, $\vec{B}^{\pm}$ is \emph{not} orthogonal to $\vec{E}^{\pm}$}.}
\end{itemize}
It is worth noticing that the absolute value of the deformed electric field and its classical counterpart is the same
\be
{\big|\vec{E}^{\pm}\big| = \big|\vec{E}^{\pm}_{\bf cl}\big|  = |k_0 \varepsilon | =: \boldsymbol{\varepsilon}}. \la{absvalsE}
\ee
The deformed dispersion relation~\eqref{defdisprel} can  also be rewritten in terms of this absolute value $\boldsymbol{\varepsilon}$,
\be
{k_0^2 - \vec{k}^2= \lambda^2 \,  \boldsymbol{\varepsilon}^2  }
\ee
By using this formula, one can easily see that,
\be
{\big|\vec{B}^{\pm}\big| = \big|\vec{B}^{\pm}_{\bf cl}\big|  = |k_0 \varepsilon | = \boldsymbol{\varepsilon}. }\la{absvalsB}
\ee
The equalities~\eqref{absvalsE} and~\eqref{absvalsE} imply that, as in the commutative case,
\be
{\big|\vec{B}^{\pm}\big| = \big|\vec{E}^{\pm}\big|.}
\ee

\section{Summary} \label{concl}
In the paper we have addressed the problem of finding the dynamical sector of Maxwell equations for Lie-Poisson noncommutativity, which be covariant and well behaved in the classical limit. We have shown that the natural equations of motion \eqn{natural} do not fulfil the requests.  

In Sec. \ref{covsec} we have found the appropriate equations of motion. Hence the full set of Maxwell equations with the right properties reads
\be
\left\{
\begin{array}{l}
   \mathcal{D}_c \,\mathcal{F}^{ca} + \frac{1}{2}\, \mathcal{F}_{de}\,f_b^{de}\, \mathcal{F}^{ab} -\mathcal{F}_{de} \, f_b^{ae}\,\mathcal{F}^{db} = 0 \\
  {\cal D}_a\,{\cal F}_{bc}-{\cal F}_{ad}\,f_b{}^{de} \,{\cal F}_{ec} \,\,\,+\,\,\,\mbox{cycl}(abc)\,\, \,\,=0.
\end{array}
\right. \la{LPM}
\ee
If an admissible Euler-Lagrange formulation is available, the dynamical equations~\eqref{LPM} are nothing but the Euler-Lagrange equations of motion, rewritten  in a gauge-covariant form.
Moreover, the left-hand side $\mathcal{E}_C^{a}$ of the dynamical equations~\eqref{LPM}  obeys the equality,
\be
\mathcal{D}_a \,\mathcal{E}^a_{C} =\frac{1}{2} \, f_{c}^{ab}\, \mathcal{F}_{ab} \,\mathcal{E}^c_{C}.
\ee
If an admissible  Lagrangian exists,  this formula is a gauge-covariant form of  Noether identity.
%\end{itemize} 

We have analysed in detail the   $\kappa$-Minkowski case and we have found a number of interesting results, which we list below.
\begin{itemize}
{\item{One may extend the covariant e.o.m. in a reasonable way, considering a one-parameter family  of  generalised e.o.m.,
\be
\mathcal{E}^a_{G}  = 0,\qquad   \mathcal{E}^a_G =  \mathcal{E}^a_C + \alpha\,\big( f_{b}^{ba}\, \mathcal{F}_{ed}\,\mathcal{F}^{ed} 
+ 4\,f_b^{be}\,\mathcal{F}_{ed}\,\mathcal{F}^{da}\big),  \qquad \alpha \in \mathbb{R},
\ee
which are compatible with the (generalised) Noether identity,
\be
\mathcal{D}_a \,\mathcal{E}^a_{G} =\frac{1}{2}\cdot\Big(1+12\,\alpha\Big) \, f_{c}^{ab}\, \mathcal{F}_{ab}   \,\mathcal{E}^c_{G}.
\ee
At $\alpha =0$ all these relations reduce to the corresponding formulae for the covariant e.o.m.
\item{At $\alpha = -1/12$ we have obtained simple solutions of the deformed e.o.m. in the vacuum case, which recover the plane waves with the left-handed and the right-handed circular polarisations at the commutative limit,
\be
{A^{\pm}_{1}  =  \varepsilon \, \sin{(k^0\,x^0 - k^3 \,x^3)},\qquad A^{\pm}_{2} = \mp \,\varepsilon \, \cos{(k^0\,x^0 - k^3 \,x^3)}  \qquad A_{0}^{\pm} = 0 =A^{\pm}_{3},}
\ee
where $k_0$ and $\vec{k}$  obey the deformed dispersion relation,
\be
{k_0^2  - \frac{\vec{k}^2}{1-\lambda^2\,\varepsilon^2}=0.}
\ee
\item The corresponding deformed electric and magnetic field read,
{\bea
\vec{E}^{\pm} &=& \varepsilon\, k^0\cdot\big(+ \cos{(k^0\,x^0 - k^3 \,x^3)}, \pm\, \sin{(k^0\,x^0 - k^3 \,x^3)},0 \big), \nonumber\\
\vec{B}^{\pm} &=&  \big( \mp\, \varepsilon\, k^3\, \sin{(k^0\,x^0 - k^3 \,x^3)}, + \varepsilon\, k^3\,\cos{(k^0\,x^0 - k^3 \,x^3)}, \pm \lambda\,\varepsilon^2\,k_0 \big). 
\eea}
The main novelty (apart from the deformed dispersion relation) is the presence of the non-zero longitudinal components of the deformed magnetic field. Therefore, in contrast with the commutative case, the fields $\vec{B}$ and $\vec{E}$ are not orthogonal to each other. However, as in the commutative case we find
\be
|\vec{B}^{\pm}| = |\vec{E}^{\pm}| = |\vec{B}^{\pm}_{\bf cl}| = |\vec{E}^{\pm}_{\bf cl}|.
\ee 
namely  their modules are equal, and coincide with the modules of the corresponding commutative limits 
$\vec{B}^{\pm}_{\bf cl}$ and $\vec{E}^{\pm}_{\bf cl}$.
}
 }
}
\end{itemize}

%%%%%
%QQQQQ
%%%%% 
\begin{appendix}%\label{appA}
% \appendixpage
%\addappheadtotoc
%\section{Appendix: }\label{appA}
\section{Proof of Proposition \ref{prop1}}\label{propa}
 \subsection*{First statement}
 Taking the partial derivative with respect to $A_j$ of the relation, 
\be
\rho\,\rho^{-1} =\mathbb{1},
\ee
 and multiplying the resulting equality by  $\rho$ from the right,
we get the identity,
\be
\partial_A^j \rho_b^m =-\rho_s^m \rho_b^q \,\partial_A^j \left(\rho^{-1}\right)^s_q.
\ee
Therefore the definition~\eqref{Lambda}  of $\mathcal{B}_b{}^{de}$ can be rewritten as follows,
\bea
\mathcal{B}_b{}^{de} &=&\left(\rho^{-1}\right)_j^d\left(-\rho_s^m \rho_b^q \,\partial_A^j \left(\rho^{-1}\right)^s_q\right)\left(\rho^{-1}\right)_m^e\,
 -(d\longleftrightarrow e) \nonumber\\
 &=& -\left(\rho^{-1}\right)_j^d  \rho_b^q \,\partial_A^j \left(\rho^{-1}\right)^e_q \,
 -(d\longleftrightarrow e),
\eea
where all the derivatives act on $\rho^{-1}$.

Using the explicit form~\eqref{universal} of the universal solution for $\rho$,
we obtain,
\be
 \left(\rho^{-1}\right)^e_q = \gamma^{e}_q - f^{te}_q A_t, 
\ee
thus,
\be
\mathcal{B}_b{}^{de} =  -\left( \gamma^{d}_j - f^{ad}_j A_a \right)  \, \left(\partial_A^j  \gamma^{e}_q - f^{je}_q  \right) \, \rho_b^q 
 -(d\longleftrightarrow e).
\ee
Expanding brackets in this expression, we get four  terms:
\be
\mathcal{B}_b{}^{de} = \big((\mathrm{I}) +(\mathrm{II}) +(\mathrm{III})  + (\mathrm{IV})\big)  \rho_b^q,
\ee
where,
\bea
(\mathrm{I}) &=&   \gamma^{e}_j \partial_A^j  \gamma^{d}_q- \gamma^{d}_j \partial_A^j  \gamma^{e}_q  , \nonumber\\
(\mathrm{II}) &=&  \gamma^{d}_j f^{je}_q  -(d\longleftrightarrow e),\nonumber \\
(\mathrm{III}) &=&  f^{ad}_j A_a \partial_A^j  \gamma^{e}_q    -(d\longleftrightarrow e),\nonumber\\
(\mathrm{IV}) &=&  f^{ae}_j A_a f^{jd}_q -f^{ad}_j A_a f^{je}_q .  
\eea
Below we demonstrate that,
\be
(\mathrm{I}) +(\mathrm{II}) +(\mathrm{III}) +(\mathrm{IV}) = f_j^{de} \left(\rho^{-1}\right)_q^j, \la{interm}
\ee
which proves the desired formula~\eqref{mainres} for the universal Poisson gauge model.

Using the first master equation~\eqref{firstsecond}, we see that,
\be
(\mathrm{I})  =  f_j^{de}\big(-\gamma^j_q\big).  \la{i1}
\ee 
The fourth term can be simplified as follows,
\bea
(\mathrm{IV}) = A_a \underbrace{\big(f^{ej}_q  f^{ad}_j + f^{dj}_q  f^{ea}_j\big)}_{-f_q^{aj} f_j^{de}} %\nonumber\\
 = -\underbrace{A_a f_q^{aj}}_{\ii \hat{A}^j_q} f_j^{de} %\nonumber\\
 = f_j^{de} \big( - \ii \hat{A}^j_q  \big), \la{i2}
\eea
where use has been made of the Jacobi identity~\eqref{Jacobi} and the definition~\eqref{AhDef} of the matrix $\hat{A}$.

The universal solution~\eqref{universal} for $\gamma$ allows to represent the third term in the following form,
\be
(\mathrm{III}) = \ii \hat{A}^{d}_j\, \partial_A^j \big[G(\hat{A})\big]^{e}_q - (d\longleftrightarrow e),
\ee
where the function $G(p)$, defined by Eq.~\eqref{GDef}, is analytic at $p=0$. Applying Proposition 4.25 of~\cite{Kupriyanov:2022ohu} at $S(p) =G(p)$,
we see that, 
\be
(\mathrm{III}) = \gamma_j^e f_q^{jd} +\gamma_q^j f_j^{de} - (d\longleftrightarrow e) = -(\mathrm{II}) + 2 f_j^{de}\gamma^j_q. \la{i3} 
\ee
Eq.~\eqref{i1} together with Eq.~\eqref{i2} and Eq.~\eqref{i3} yield,
\be
(\mathrm{I}) +(\mathrm{II}) +(\mathrm{III}) +(\mathrm{IV}) = f_j^{de} \big( \gamma - \ii \hat{A}\big)_q^j. 
\ee
Confronting this relation with the explicit expression~\eqref{universal} for $\rho$, we immediately arrive at Eq.~\eqref{interm}, which completes
the  proof of Eq.~\eqref{mainres} for the universal Poisson gauge theory.  \\

\subsection*{Second statement}
We now  prove the validity of Eq.~\eqref{mainres} for any Poisson gauge model, which is related to the universal one through the Seiberg-Witten map, discussed in the Introduction.
Let the matrix $\tilde\rho(\tilde A)$ be related to the universal  solution $\rho(A)$ via Eq.~\eqref{redef} and Eq.~\eqref{newGR}. 
Introducing the notation,
\be
\tilde{\partial}_A^a := \frac{\partial}{\partial \tilde{A}_a},
\ee 
one can easily check that the structure~\eqref{Lambda} remains unchanged upon the Seiberg-Witten maps:
\bea
\tilde{\mathcal{B}}_b{}^{de}(\tilde{A}) &:=& \left(\tilde\rho^{-1}(\tilde{A})\right)_j^d\left(\tilde\partial^j_A\tilde\rho_b^m(\tilde{A})-\tilde\partial^m_A\tilde{\rho}_b^j(\tilde{A})\right)\left(\tilde{\rho}^{-1}(\tilde{A})\right)_m^e \nonumber\\
&=& \bigg[\frac{\partial \tilde A_j}{\partial A_p}  \left(\rho^{-1}(A)\right)_p^d \frac{\partial \tilde A_m}{\partial A_r}  \left(\rho^{-1}(A)\right)_r^e \bigg]\bigg|_{A = A(\tilde{A})}\, \tilde{\partial}_A^j\bigg(\frac{\partial A_s}{\partial \tilde{A}_m}\Big[\rho_b^s(A)\Big]\Big|_{A = A(\tilde{A})}\bigg) - (d\longleftrightarrow e) \nonumber\\
&=&  \bigg[ \left(\rho^{-1}(A)\right)_p^d \underbrace{ \frac{\partial \tilde A_m}{\partial A_r}\frac{\partial A_s}{\partial \tilde{A}_m} }_{\delta_s^r} \left(\rho^{-1}(A)\right)_r^e \bigg]\bigg|_{A = A(\tilde{A})}\, \underbrace{\frac{\partial \tilde A_j}{\partial A_p}\bigg|_{A = A(\tilde{A})} \tilde{\partial}_A^j\bigg(\Big[\rho_b^s(A)\Big]\Big|_{A = A(\tilde{A})}\bigg) }_{\big[\partial_A^p \rho^s_b(A)\big]\big|_{A=A(\tilde{A})}} \nonumber\\
&-& (d\longleftrightarrow e) \nonumber\\
&+&  \underbrace{\bigg[\bigg(\frac{\partial \tilde A_j}{\partial A_p}  \left(\rho^{-1}(A)\right)_p^d \frac{\partial \tilde A_m}{\partial A_r}  \left(\rho^{-1}(A)\right)_r^e \rho_b^s(A)\bigg) - (d\longleftrightarrow e) \bigg]\bigg|_{A = A(\tilde{A})}}_{\mbox{\tiny{skew-sym. in ($j$,$m$)}} }\cdot\underbrace{\frac{\partial^2 A_s}{\partial \tilde{A}_j \partial\tilde{A}_m} }_{\mbox{\tiny{sym. in ($j$,$m$)}} } 
\nonumber\\
&=& \bigg[ \left(\rho^{-1}(A)\right)_p^d 
\left(\partial_A^p \rho^s_b(A) -\partial_A^s \rho^p_b(A)\right)
\left(\rho^{-1}(A)\right)_s^e \bigg]\bigg|_{A = A(\tilde{A})} \nonumber\\
&=&\bigg[ {\mathcal{B}}_b{}^{de}({A}) \bigg]\bigg|_{A = A(\tilde{A})}. \la{struinv}
\eea 
As we have proven above, in the universal Poisson gauge theory, ${\mathcal{B}}_b{}^{de}({A}) = f_b^{de}$. Therefore, Eq.~\eqref{struinv} yields, 
$
\tilde{\mathcal{B}}_b{}^{de}(\tilde{A})  = f_b^{de}.
$
Q.E.D.

\section{Derivation of Eqs.~\eqref{PCstep1}, \eqref{PCstep2}}\label{appbc}
This derivation of Eq. \eqref{PCstep1} is quite short.
\bea
\left\{A_a,\rho_b^a\right\} &=& \partial_A^s\rho_b^a \, \left\{A_a,A_s\right\}\nonumber\\
&=&\frac{1}{2} \, \big(\partial_A^s \rho_b^a -\partial_A^a \rho_b^s\big)\,  \left\{A_a,A_s\right\} \nonumber\\
&=& - \frac{1}{2} \, \big(\partial_A^s \rho_b^a -\partial_A^a \rho_b^s\big)\cdot \underbrace{\delta_s^q}_{(\rho^{-1})_s^r\rho^q_r}\cdot\underbrace{\delta_a^l}_{(\rho^{-1})^t_a \rho_t^l}\cdot \left\{A_q,A_l\right\} \nonumber\\
&=& - \frac{1}{2}\underbrace{\left(\rho^{-1}\right)_s^r\,\big(\partial_A^s \rho_b^a -\partial_A^a \rho_b^s\big)\,\left(\rho^{-1}\right)^t_a }_{f_b^{rt}} 
\cdot
\rho^q_r\, \rho_t^l\, \left\{A_q,A_l\right\}
\nonumber\\
&=& - \frac{1}{2}\,f_b^{rt}
\,
\rho^q_r\, \rho_t^l\, \left\{A_q,A_l\right\},
\eea
where at the last step we used Proposition~\ref{prop1}. Q.E.D.

 The derivation of Eq.~\eqref{PCstep2} is more sophisticated than the previous one. We first derive it  for the universal Poisson gauge model. 

We start from the main formula~\eqref{mainres} of Proposition \ref{prop1}, representing it as follows, 
\be
\partial_A^s\rho_b^r - \partial_A^r\rho_b^s = f_b^{lq}\, \rho_l^s\,\rho_q^r.
\ee 
Contracting this equation with the combination $f_r^{pj} A_j$, and renaming the mute indices in a suitable way, we obtain:
\be
f_q^{pl}\, A_l\,\partial_A^s\rho_b^q - f_r^{pj}\, A_j\, \partial_A^r\rho_b^s = A_j \,f_r^{jp}\,f_b^{ql} \,\rho_q^r\,\rho_l^s. \la{stepi2}
\ee
According to the structure~\eqref{universal} of the universal solutions, $\rho(A) = F\big(\hat{A}\big)$, 
with $F(p)$ being an analytic function at $p=0$. Therefore the second term of the left-hand side can be rewritten as follows,
 \bea
 - f_r^{pj}\, A_j\, \partial_A^r\rho_b^s &=& \ii  \, \hat{A}^{p}_r \,\partial_A^r [F\big(\hat{A}\big)]^{s}_b \nonumber\\
 &=& [F\big(\hat{A}\big)]^s_l \,f_b^{lp} +[F\big(\hat{A}\big)]^q_b \,f^{ps}_q \nonumber\\
 &=& f^{ps}_q \,\rho^{q}_b - f^{pl}_{b}\,\rho^{s}_l, \la{secondTermSimplif}
 \eea
where use has been made of Proposition~4.25 of Ref.~\cite{Kupriyanov:2022ohu}.

%The relations~\eqref{stepi2} and~\eqref{secondTermSimplif} yield,
Substituting this result in Eq.~\eqref{stepi2}, and contracting the resulting equality with $-\partial_p A_s$, we get, 
\be
-\big(f_q^{pl} \,A_l \,\partial_A^{s}\rho_b^q  + f_q^{ps} \rho_b^{q} \big) \partial_p A_s=  -\big(f^{pl}_{b}\,\rho^{s}_l + A_j \,f_r^{jp}\,f_b^{ql} \,\rho_q^r\,\rho_l^s\big) \partial_p A_s. \la{stepi3}
\ee
Below we demonstrate that the lhs and rhs of this relation respectively reproduce the lhs and rhs of the required formula~\eqref{PCstep2}. Indeed, the lhs of Eq.~\eqref{stepi3} can be elaborated as follows,
\bea
\mathbf{l.h.s.} &=&- \big(f_q^{pl} \,A_l \,\partial_A^{s}\rho_b^q  + f_q^{ps} \rho_b^{q} \big) \partial_p A_s  = -\partial_A^s\big(f_q^{pl} \,A_l \,\rho_b^q  \big) \partial_p A_s =- \partial_p \big(\underbrace{f_q^{pl} \,A_l}_{-\ii \hat{A}_q^p} \,{\rho_b^q}  \big) \nonumber\\
&=& \ii \, \partial_p \big(\hat{A} \,\rho\big)^{p}_b =  \partial_p \big(\underbrace{\mathbb{1}  + \ii \,\hat{A} \,\rho }_{\gamma\rho,\,\,\mbox{\scriptsize{see Eq.~\eqref{universal}}}}\big)^{p}_b =
\partial_p( \gamma^p_q\rho^q_b),
\eea
which is nothing but the lhs of Eq.~\eqref{PCstep2}. Now we take a closer look at the rhs of Eq.~\eqref{stepi3}:
\bea
\mathbf{r.h.s.} &=&-\big(f^{pl}_{b}\,\rho^{s}_l + A_j \,f_r^{jp}\,f_b^{ql} \,\rho_q^r\,\rho_l^s\big) \partial_p A_s = 
-f_b^{ql}\,\big(\delta_q^p + \underbrace{f_r^{jp}\,A_j}_{\ii\hat{A}^p_r}\,\rho_q^r\big)\,\rho_l^s\, \partial_p A_s \\
&=&-f_b^{ql}\,\big(\underbrace{\mathbb{1} + \ii\,\hat{A}\,\rho}_{\gamma\rho}\big)^p_q\,\rho_l^s\, \partial_p A_s
=-f_b^{ql}\,\gamma_r^p\,\rho^r_q\,\rho^{s}_l\,  \partial_p A_s ,\nonumber\\
&=& -\frac{1}{2}\,f_b^{ql}\,\rho^r_q\,\rho^{s}_l\big(\gamma_r^p\,\partial_p A_s -\gamma_s^p\,\partial_p A_r \big),
\eea
where one can easily recognize the rhs of the desired equation~\eqref{PCstep2}. Summarising, we have proven Eq.~\eqref{PCstep2} for the universal Poisson gauge model.

If a given Poisson model, defined by the matrices $\tilde\gamma(\tilde{A})$ and $\tilde{\rho}(\tilde{A})$, is related to the universal one through the Seiberg-Witten map, i.e. through the invertible field redefinition~\eqref{redef}, accompanied by the relations~\eqref{newGR}, then the equality~\eqref{PCstep2} takes place as well. Indeed, performing a few simple calculations one can easily check that,
\bea
&&\partial_l (\tilde\gamma_a^l(\tilde{A}) \tilde\rho_b^a(\tilde{A})) + \frac{1}{2}\, f_b^{rt}\,\tilde{ \rho}_r^q(\tilde{A})\,\tilde{\rho}_t^l(\tilde{A})\,(\tilde{\gamma}(\tilde{A})_q^p\,\partial_p \tilde{A}_l - \tilde{\gamma}_l^p(\tilde{A})\,\partial_p \tilde{A}_q)\nonumber\\
&&=\big[\partial_l (\gamma_a^l({A}) \rho_b^a({A})) + \frac{1}{2}\, f_b^{rt}\,{ \rho}_r^q({A})\,{\rho}_t^l({A})\,({\gamma}({A})_q^p\,\partial_p {A}_l - {\gamma}_l^p({A})\,\partial_p {A}_q)\big]_{A = A(\tilde{A})},
\eea
where $\gamma(A)$ and $\rho(A)$ denote the universal solutions~\eqref{universal} of the master equations. Since, Eq.~\eqref{PCstep2} is valid for the universal Poisson gauge model, the expression inside the square brackets vanishes. Therefore,
\be
\partial_l (\tilde\gamma_a^l(\tilde{A}) \tilde\rho_b^a(\tilde{A})) =- \frac{1}{2}\, f_b^{rt}\,\tilde{ \rho}_r^q(\tilde{A})\,\tilde{\rho}_t^l(\tilde{A})\,(\tilde{\gamma}(\tilde{A})_q^p\,\partial_p \tilde{A}_l - \tilde{\gamma}_l^p(\tilde{A})\,\partial_p \tilde{A}_q),
\ee
Q.E.D.

\section{Violation  of Eq.~\eqref{mucommlim} in the $\kappa$-Minkowski case}\label{appd}
From now on we assume that the measure $\mu(x,\lambda)$  exhibits a finite limit  at $\lambda\to 0$, because otherwise it would not make sense to consider Eq.~\eqref{mucommlim} at all.

Upon rescaling of  coordinates, 
 the general solution~\eqref{muhigher} rescales as follows,
\be
\mu(\Omega x,\lambda) = \Omega^{1-d}\,\mu(x,\lambda), \qquad \forall \lambda, \quad \forall\Omega\in\mathbb{R}. \la{tpmu0}
\ee
  Since the identity~\eqref{tpmu} is valid at any $\lambda$, and, by assumption, $\lim_{\lambda\to 0}\mu(x,\lambda)$ exists, this identity is valid
at the commutative limit as well,
 \be
 \lim_{\lambda\to 0}\mu(\Omega x,\lambda) = \Omega^{1-d}\,\lim_{\lambda\to0}\mu(x,\lambda), \qquad \forall\Omega\in\mathbb{R}. \la{tpmu}
 \ee
 If the relation~\eqref{mucommlim} was possible for some “lucky" choice of the arbitrary function $F_{\lambda}$, the formula~\eqref{tpmu} would yield the controversary equality,
 \be
 1=\Omega^{1-d}, \qquad \forall\Omega\in\mathbb{R}.
 \ee 
Thus we conclude that the $\kappa$-Minkowski non-commutativity is not compatible with the limit~\eqref{mucommlim}. Q.E.D.

\section{Proof of Proposition \ref{propD}}\label{appe}
On substituting the explicit form~\eqref{CovEOM} of $\mathcal{E}^a_{C}$ in the l.h.s of the desired equality~\eqref{usfulidentity}, one obtains, 
\be
\mathcal{D}_a \,\mathcal{E}^a_{C} 
%= \mathcal{D}_a\mathcal{D}_c \mathcal{F}^{ca} + \frac{1}{2}\,\mathcal{D}_a \mathcal{F}_{de}\,f_b^{de}\, \mathcal{F}^{ab}
%+ \frac{1}{2}\, \mathcal{F}_{de}\,f_b^{de}\,\mathcal{D}_a \mathcal{F}^{ab}
 %-\mathcal{D}_a\mathcal{F}_{de} \, f_b^{ae}\,\mathcal{F}^{db}
 % -\mathcal{F}_{de} \, f_b^{ae}\,\mathcal{D}_a\mathcal{F}^{db} \nonumber\\
  = \frac{1}{2}[\mathcal{D}_a,\mathcal{D}_c] \mathcal{F}^{ca} + \frac{1}{2}\,\mathcal{D}_a \mathcal{F}_{de}\,f_b^{de}\, \mathcal{F}^{ab}
+ \frac{1}{2}\, \mathcal{F}_{de}\,f_b^{de}\,\mathcal{D}_a \mathcal{F}^{ab}
 -\mathcal{D}_a\mathcal{F}_{de} \, f_b^{ae}\,\mathcal{F}^{db}
  -\mathcal{F}_{de} \, f_b^{ae}\,\mathcal{D}_a\mathcal{F}^{db}
\ee
Applying the identity~\eqref{comD}, we see that 
\be
\frac{1}{2}[\mathcal{D}_a,\mathcal{D}_c] \mathcal{F}^{ca} = \mathcal{F}_{de} \, f_b^{ae}\,\mathcal{D}_a\mathcal{F}^{db},
\ee
therefore,
\be
\mathcal{D}_a \,\mathcal{E}^a_{C}  =    \frac{1}{2}\,\mathcal{D}_a \mathcal{F}_{de}\,f_b^{de}\, \mathcal{F}^{ab}
+ \frac{1}{2}\, \mathcal{F}_{de}\,f_b^{de}\,\mathcal{D}_a \mathcal{F}^{ab}
 -\mathcal{D}_a\mathcal{F}_{de} \, f_b^{ae}\,\mathcal{F}^{db}. 
\ee
Renaming the mute indices appropriately, one can easily get
\bea
-\mathcal{D}_a\mathcal{F}_{de} \, f_b^{ae}\,\mathcal{F}^{db} = 
\mathcal{D}_a\mathcal{F}_{bd} \,f^{ab}_e\,\mathcal{F}^{de} = \mathcal{D}_b\mathcal{F}_{da} \,f^{ab}_e\,\mathcal{F}^{de}, \nonumber\\
\Longrightarrow\quad -\mathcal{D}_a\mathcal{F}_{de} \, f_b^{ae}\,\mathcal{F}^{db} = \frac{1}{2}\,f^{ab}_e\,\mathcal{F}^{de}\big(\mathcal{D}_a\mathcal{F}_{bd}   + \mathcal{D}_b\mathcal{F}_{da}  \big),
\eea
and also,
\be
\frac{1}{2}\,\mathcal{D}_a \mathcal{F}_{de}\,f_b^{de}\, \mathcal{F}^{ab} = \frac{1}{2}\,f^{ab}_e\,\mathcal{F}^{de}\,\mathcal{D}_d\mathcal{F}_{ab},  
\ee
thus,
\bea
\mathcal{D}_a \,\mathcal{E}^a_{C}  &=&   
 \frac{1}{2}\, \mathcal{F}_{de}\,f_b^{de}\cdot\underbrace{\mathcal{D}_a \mathcal{F}^{ab}}_{
 \mathcal{E}_C^b - \frac{1}{2}\,\mathcal{F}_{cp}\,f_a^{cp}\,\mathcal{F}^{ba} +\mathcal{F}_{cp}\,f_{a}^{bp}\, \mathcal{F}^{ca}
 } 
 +\frac{1}{2}\,f^{ab}_e\,\mathcal{F}^{de}\big(\underbrace{\mathcal{D}_a\mathcal{F}_{bd}   + \mathrm{cycl}(a\,b\,d)}_{
 \mathcal{F}_{ac}\,f_{b}^{cp}\,\mathcal{F}_{pd} + \mathrm{cycl}(a\,b\,d)
 } \big)\nonumber\\
 &=&  \frac{1}{2}\, \mathcal{F}_{de}\,f_b^{de}\,\mathcal{E}_C^b + \mathcal{Q}, \la{propDinterm}
\eea
where the first term is nothing but the desired contribution, whilst
\be
\mathcal{Q}  = (\mathrm{I}) + (\mathrm{II}) +(\mathrm{III}) + (\mathrm{IV}) +(\mathrm{V}),
\ee
with
\bea
(\mathrm{I}) &=& - \frac{1}{4}\, \mathcal{F}_{de}\,f_b^{de}\,\mathcal{F}_{cp}\,f_a^{cp}\,\mathcal{F}^{ba},\nonumber\\
(\mathrm{II})&=&  +\frac{1}{2}\, \mathcal{F}_{de}\,f_b^{de}\,\mathcal{F}_{cp}\,f_{a}^{bp}\, \mathcal{F}^{ca}, \nonumber\\
(\mathrm{III})&=&+\frac{1}{2}\,f^{ab}_e\,\mathcal{F}^{de}\,   \mathcal{F}_{ac}\,f_{b}^{cp}\,\mathcal{F}_{pd}, \nonumber\\
(\mathrm{IV})&=&+\frac{1}{2}\,f^{ab}_e\,\mathcal{F}^{de}\,   \mathcal{F}_{bc}\,f_{d}^{cp}\,\mathcal{F}_{pa}, \nonumber\\
(\mathrm{V})&=& +\frac{1}{2}\,f^{ab}_e\,\mathcal{F}^{de}\,  \mathcal{F}_{dc}\,f_{a}^{cp}\,\mathcal{F}_{pb}. 
\eea
 In order to obtain Eq.~\eqref{propDinterm}, we used the deformed Bianchi identity~\eqref{Bianchi} and the definition~\eqref{CovEOM} of the covariant e.o.m. Below we demonstrate that $\mathcal{Q}=0$, what will complete our proof.
 
The contributions $(\mathrm{I})$ and $(\mathrm{IV})$, being contractions of symmetric and antisymmetric tensors, vanish identically:
\bea
(\mathrm{I}) &=& - \frac{1}{4}\cdot \underbrace{\mathcal{F}_{de}\,f_b^{de}\,\mathcal{F}_{cp}\,f_a^{cp}}_{\mbox{\scriptsize{sym. in $(a,b)$}}}\cdot\underbrace{\mathcal{F}^{ba}}_{\mbox{\scriptsize{skew-sym. in $(a,b)$}}} = 0, \nonumber\\
(\mathrm{IV})&=&+\frac{1}{2}\cdot\underbrace{\mathcal{F}^{de}}_{\mbox{\scriptsize{skew-sym. in $(d,e)$}}}\cdot \underbrace{ f^{ab}_e\,  \mathcal{F}_{bc}\,f_{d}^{cp}\,\mathcal{F}_{pa}}_{\mbox{\scriptsize{sym. in $(d,e)$}}} = 0.
\eea 
Renaming  the mute indices, we see that the sum of the remaining three contributions equals to zero, 
\be
(\mathrm{II}) + (\mathrm{III}) +(\mathrm{V}) = \frac{1}{2}\,\mathcal{F}^{ca}\,\mathcal{F}_{de}\,\mathcal{F}_{cp}\,
 \underbrace{\big(f^{de}_b f^{bp}_a + f^{ep}_b f^{bd}_a + f^{pd}_b f^{be}_a\big)}_0 = 0,
\ee
where use has been made of the Jacobi identity~\eqref{Jacobi}.  Q.E.D.

\section{$\kappa$-Minkowski non-commutativity: new and old solutions }\label{appf}
In the forthcoming discussion we use the notations,
\bea
g(z) := \sqrt{z^2+1} + z,\qquad g^{\prime}(z)  \equiv \frac{\dd g(z)}{\dd z} = \frac{ \sqrt{z^2+1} + z}{ \sqrt{z^2+1} }.
\eea
For the $\kappa$-Minkowski non-commutativity, the first matrix $\gamma$ viz, 
\bea
\tilde\gamma(A)  &=&  \left(
\begin{array}{cccc}
g\left(\frac{\lambda A_0}{2}\right)-\frac{\lambda A_0}{2} &-\frac{\lambda A_1}{2} &-\frac{\lambda A_2}{3} & -\frac{\lambda A_3}{2} \\
0 &g\left(\frac{\lambda A_0}{2}\right) &0 &0 \\
0 &0 &g\left(\frac{\lambda A_0}{2}\right) &0 \\
0 &0 &0 &g\left(\frac{\lambda A_0}{2}\right) 
\end{array}
\right),  
\eea
was obtained in~\cite{Kupriyanov:2020axe}.
The corresponding matrix $\rho$, namely,
\bea
\tilde\rho(A)  &=&  \left(
\begin{array}{cccc}
\frac{1}{\sqrt{ \left(\frac{\lambda A_0}{2}\right)^2+1}} &-g^{\prime}\left(\frac{\lambda A_0}{2}\right)\frac{\lambda A_1}{2} &-g^{\prime}\left(\frac{\lambda A_0}{2}\right)\frac{\lambda A_2}{3} & -g^{\prime}\left(\frac{\lambda A_0}{2}\right)\frac{\lambda A_3}{2} \\
0 &g\left(\frac{\lambda A_0}{2}\right) &0 &0 \\
0 &0 &g\left(\frac{\lambda A_0}{2}\right) &0 \\
0 &0 &0 &g\left(\frac{\lambda A_0}{2}\right) 
\end{array}
\right),
\eea
was found in~\cite{Kupriyanov:2021aet}. The Seiberg-Witten map, which connects this first Poisson gauge model with the universal one, has been constructed in~\cite{Kupriyanov:2022ohu}. One has to set $a^0 = \lambda/2$, $a^1=0$, $a^2 = 0$, ..., $a^{d-1} =0$, in the corresponding formulae of the quoted references. We emphasise that in the present project we are \emph{not} working with the generalised $\kappa$-Minkowski non-commutativity, restricting ourselves to the standard one only.  

The Poisson gauge theory, based on the novel simple matrices~\eqref{newRG}, is related to the model, defined by  $\tilde{\gamma}$ and $\tilde{\rho}$,
through the Seiberg-Witten map as well. Indeed, one can easily check that upon the invertible fields redefinition,
\be
\left\{ 
\begin{array}{l}
 A_0 = \frac{2}{\lambda}\, \mathrm{arcsinh}\left(\frac{\lambda \tilde{A}_0}{2}\right), \\
 A_b = \Big(g\big(\frac{\lambda \tilde{A}_0}{2}\big)\Big)^{-1}\cdot \tilde{A}_b, \,\,b=1,2,3, 
\end{array}
\right.
\quad\Longleftrightarrow\quad
\left\{ 
\begin{array}{l}
 \tilde{A}_0 = \frac{2}{\lambda}\, \mathrm{sinh}\left(\frac{\lambda {A}_0}{2}\right), \\
 \tilde{A}_b = e^{\frac{\lambda {A}_0}{2}}\cdot {A}_b, \,\,b=1,2,3, 
\end{array}
\right.
\ee
the relations~\eqref{newGR} are satisfied. Hence, by transitivity,  there exists a Seiberg-Witten map between the universal Poisson gauge model and the  one, introduced in Sec.~\ref{KappaES}.
%%%%%
%QQQQQ
%%%%%
\end{appendix}
%\end{document}

{\section*{Acknowledgements} M.K. and P.V. acknowledge partial financial support from INFN through Iniziativa Specifica GEOSYM-QFT. The research of P.V. was carried out in the frame of Programme STAR Plus,
    financially supported  by UniNA and Compagnia di San Paolo. V.G.K. acknowledges support from the CNPq Grant 304130/2021-4 and the FAPESP Grant 2021/09313-8. }

\end{document}